\documentclass[10pt, conference]{IEEEtran}
\IEEEoverridecommandlockouts
\usepackage{cite}
\usepackage{amsmath,amssymb,amsfonts}
\usepackage{algorithmic}
\usepackage{graphicx}
\usepackage{textcomp}
\usepackage{xcolor}
\usepackage{url}
\usepackage[bookmarks=false]{hyperref}
\usepackage{graphicx,alltt,xcolor,float}
\usepackage{listings}
\usepackage{comment}
\usepackage[shortlabels]{enumitem}
\usepackage{listings}
\usepackage[utf8x]{inputenc} 
\usepackage{multirow}
\usepackage{amsmath}

\def\BibTeX{{\rm B\kern-.05em{\sc i\kern-.025em b}\kern-.08em
    T\kern-.1667em\lower.7ex\hbox{E}\kern-.125emX}}

\begin{document}

\title{Vulnerability Assessment and Penetration Testing on IP cameras}

\author{
\IEEEauthorblockN{Pietro Biondi, Stefano Bognanni and Giampaolo Bella}
\IEEEauthorblockA{\textit{Dipartimento di Matematica e Informatica}\\
\textit{Universit\`a di Catania}\\
Catania, Italy\\
pietro.biondi@phd.unict.it, stefano.bognanni97@gmail.com, giamp@dmi.unict.it}
}

\maketitle

\begin{abstract}
IP cameras have always been part of the Internet of Things (IoT) and are among the most widely used devices in both home and professional environments. 
Unfortunately, the vulnerabilities of IP cameras have attracted malicious activities. For example, in 2016, a massive attack resulted in thousands of cameras and IoT devices being breached and used to create a botnet.
Given this history and the extremely sensitive nature of the data these devices have access to, it is natural to question what security measures are in place today.

In this paper, a vulnerability assessment and penetration testing is performed on a specific model of IP camera, the TP-Link Tapo C200. More in detail, our findings show that the IP camera in question suffers from three vulnerabilities such as: denial of service, video eavesdropping and, finally, a new type of attack called \lq\lq Motion Oracle".
Experiments are not limited to the offensive part but also propose countermeasures for the camera in question and for all those that may suffer from the same vulnerabilities. The countermeasure is based on the use of another IoT device, a Raspberry Pi.

\end{abstract}
\begin{IEEEkeywords}
IoT, VAPT, attacks, profiling, cybersecurity, privacy
\end{IEEEkeywords}
\section{Introduction}\label{sec:intro}
With the explosion of the Internet of Things (IoT) world, many devices already on the market have undergone a transformation, becoming more functional and cost-effective.
One of these is undoubtedly the IP camera, a video surveillance camera that uses the network to transmit audio and video signals. Unfortunately, IP cameras, like many other IoT devices, are the target of countless malicious cyber attacks that often aim to violate the confidentiality of transmitted data.

Today, IP cameras are being used to make CCTV, as confirmed by a 2019 estimate of around 770 million IP cameras being used in this field~\cite{survcamera}.
In addition, IP cameras are increasingly being used in peoples' houses. Thanks to their low cost, they are used, for example, to monitor a house when people are away, to monitor pets and also as baby monitors.
In view of the main uses, it is clear that the nature of the data transmitted by these devices is decidedly intimate and private.
Therefore, it would seem reasonable to assume that a series of countermeasures are implemented to protect the confidentiality of the transmissions.
However, there are no general guidelines for the realisation of such products and often the precautions taken are very weak or even completely absent.

The security of the IoT world has been at the centre of the global debate for a few years now. In fact, in October 2016, the Mirai malware was able to build a botnet, compromising thousands of devices, including many IP cameras~\cite{mirai}. In 2017, the list of known vulnerabilities within the CCTV Calculator~\cite{cctvcalc} was published, involving IP cameras from many vendors, including Samsung, D-link, Panasonic, Cisco, Hikvision, FOSCAM, Y-Cam, TP-Link, AirLive, Sony, Arecont Vision,  Linksys, Canon and many more.
The topic of security and IP cameras is an open one, so in this paper we decided to carry out a vulnerability assessment and penetration testing process on one of today's IP cameras in particular, the TP-Link Tapo C200~\cite{c200}.

\subsection{Contributions}
This article explores whether and how IP cameras can be exploited today using freeware, i.e. non-commercial tools. We address this question in a scenario where there is an insider attacker who connects to the same network as the IP camera with her attacking laptop via Ethernet or Wi-Fi connectivity.
The results are that, relying on simple tools such as nmap, Ettercap, Wireshark and Python programming, the attacker can seriously compromise the service and undermine user privacy.

More precisely, we describe all our results obtained through the reverse engineering process performed on the Tapo camera; thanks to this process we were able to perform our vulnerability assessment.

The vulnerability assessment exercise allowed us to obtain information and evaluating a ceremony for everything related to proprietary services and third-party video streams. The work continued with penetration testing, which aim to exploit vulnerabilities obtained from the vulnerability assessment exercise on the Tapo camera. Specifically, we found that the Tapo camera suffers from: Denial of Service which clearly renders the camera unavailable; Video eavesdropping which affects user privacy; in addition, we demonstrate a third attack called \lq\lq Motion Oracle" which allows an attacker on the same network as the Tapo C200 to detect traffic related to the camera's motion detection service and use it to obtain information about the victim by analysing the frequency of generated traffic.

Finally, we prototype effective countermeasures, at least against video eavesdropping, by taking advantage of a cheap Raspberry Pi~\cite{raspb} device. This can implement encryption so that packets are only transferred in encrypted form over the network and therefore cannot be understood by a man in the middle (MITM).
The Raspberry Pi then transmits the encrypted packets from the IP Camera, ensuring user security, in a completely transparent manner.

\subsection{Testbed}
Our testbed consists of some useful devices for VAPT experiments.

The Tapo C200 is an entry-level IP camera designed for home use. Released in 2020 with a very low cost (around 30€), the Tapo C200 has several technical features, thanks to which it boasts several functionalities, such as: compatibility with the main smart home systems (e.g. Alexa~\cite{alexa} and Google Assistant~\cite{GoogleAssistant}); night vision, resolution up to Full HD; motion detection; acoustic alarm; data storage; voice control; use with third-party software; webcam mode.

To use the Tapo C200, a TP-Link account is required which allows access to the various cloud services offered.
Access to the TP-Link account is via the Tapo app, which is available free of charge on Google Play~\cite{tapoGPlay} and APP store~\cite{tapoApple} for devices running Android 4.4+ and IOS 9+ respectively. The application has a simple and intuitive user interface, from which it is possible to use and manage the devices of the Tapo family, including the Tapo C200. The range of services offered for the Tapo C200 includes: Remote access and control of the Tapo C200; Sharing the Tapo C200 with other accounts; synchronisation of settings; integrating the Tapo C200 with smart home systems; and receiving motion notifications.

\begin{figure}[ht]
 \centering
 \includegraphics[scale=0.35]{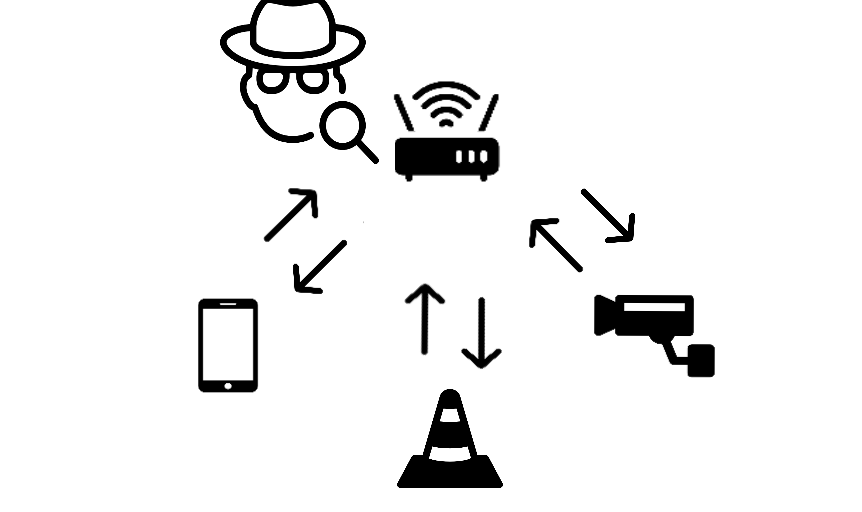}
 \caption{Original testbed}
 \label{fig:testbed}

\end{figure}

Below, the Figure~\ref{fig:testbed} shows the testbed on which the experiments were conducted and how the elements of the testbed are connected to each other, note that all connections are made via a Wi-Fi network.
In a nutshell, the testbed consists of:

\begin{enumerate}[(a)]

\item A Wi-Fi switch to which the various devices were connected
\item The Tapo C200 IP Camera
\item A smartphone on which SSL Packet Capture and the Tapo application were installed
\item A Linux machine on which Ettercap and Wireshark tools have been installed
\item A Linux machine on which the third-party players iSpy and VLC have been installed

\end{enumerate}

By monitoring the environment with the various instruments, the Tapo C200 was used as a normal user, with the aim of extracting useful information for operational analysis.

\subsection{Related Work}\label{sec:RW}
As early as 2015, it was shown that IP cameras can encompass multiple research areas, including multimedia security, network security and cloud security, and end-user privacy.
In fact, Tekeoglu and Tosun looked into the security of cloud-based wireless IP cameras~\cite{cloudbasedIPcamerajpeg}. They studied the traffic generated by a wireless IP camera that is easy to set up for the average home user. Hence, they proved that an attacker can sniff the IP camera's network traffic and would be able to reconstruct the JPEG images from the data stream. Based on their information, their system has some limitations, e.g. their script only reconstructed 253 JPEG images over about 20 hours of video track.

A few years later in 2017, Liranzo and Hayajneh presented an analysis of the main privacy and security issues affecting thousands of IP camera consumers globally, they proposed a series of recommendations to help protect consumers' IoT devices in the intimacy of their home~\cite{2017issueIPcamera}.

It is clear that modern IP cameras collect a large amount of data, which is then transferred to cloud systems. 
As a result, IP cameras process data to such an extent that it becomes important to assess the security of these cameras also from the point of view of privacy for users.

\subsection{Paper Structure}

This manuscript continues to describe all background information useful for understanding the content of the paper (\S\ref{sec:background}). The core of the paper begins with the reverse engineering process (\S\ref{sec:reverse}) which continues with the vulnerability assessment (\S\ref{sec:VA}) and the penetration testing (\S\ref{sec:PT}) performed on the Tapo C200 IP camera. Finally, the paper describes a countermeasure (\S\ref{sec:countermeasures})
and concludes with some broader evaluations of the results (\S\ref{sec:conclusions}).

\section{Background}~\label{sec:background}
In this section we provide all the background information necessary for understanding all the steps performed in the VAPT work on the Tapo C200.
The IP camera also has the functionality to integrate the device with third-party infrastructures and systems. Such systems require the use of dedicated protocols such as the Real Time Streaming Protocol (RTSP)~\cite{rtsp}.

RTSP is a network protocol used for multimedia streaming systems, its purpose is to orchestrate the exchange of media, such as audio and video, across the network. RTSP extends the RTP~\cite{rtp} and RTCP~\cite{rtcp} protocols by adding the necessary directives for streaming, such as:
\begin{itemize}
 \item Describe: is used by the client to obtain information about the desired resource
 \item Setup: specifies how a single media stream is to be transported
 \item Play/Pause: starts/stops multimedia playback
 \item Record: used to store a stream
\end{itemize}

ONVIF (Open Network Video Interface Forum) is an organisation whose aim is to promote compatibility between video surveillance equipment so that systems made by different companies are interoperable.
ONVIF provides an interface for communication with the different equipment through the use of various protocols. ONVIF provides different configuration profiles to make the best use of the different technical features.

Profiles G, Q, S and T, are dedicated to video systems, however, the Tapo C200 is only compatible with Profile S~\cite{comp}.
Profile S is one of the most basic profiles and is designed for IP-based video systems, the features provided for Profile S that are compatible with the Tapo C200 are user authentication, NTP support and H264~\cite{onvifs} audio and video streaming using RTSP.

In order to take advantage of the multimedia streaming services, it was necessary to use the open source, multi-platform media player VLC~\cite{vlc} and the iSpy DVR~\cite{ispy}.
In order to scan the services exposed by the device, the software Nmap~\cite{nmap} was used, which is able to perform port scanning operations, and the software Nessus~\cite{nessus}, which allows to perform a risk analysis on network devices, indicating the criticality level of vulnerabilities according to the CVSS standard. 
CVSS is an industry standard designed to assess the severity of computer system security vulnerabilities~\cite{cvss}. The assessment consists of a score on a scale from 0 to 10 calculated on the basis of a formula~\cite{scoreCvssdoc} that takes into account Attack Vector (AV), Scope (S), Attack Complexity (AC), Confidentiality (C), Privileges Required (PR) \& Integrity (I), User Interaction (UI) and Availability (A). The higher the score obtained based on the values assigned to these metrics, the more serious the vulnerability.

Finally, the Tapo C200 was analysed using the tools Ettercap~\cite{ettercap} to intercept traffic to and from the device, SSL packet capture~\cite{sslpacket} to intercept and read traffic to and from the Tapo application and Wireshark~\cite{wireshark} to understand network traffic and use the H264 extractor extension to convert network traffic into playable video.

\section{Reverse engineering}~\label{sec:reverse}

Our work starts from the reverse engineering process that allowed us to deduce that: when the application and the camera are on different networks, a classic client-server architecture is used in which the Tapo Server is used to distribute information, such as the device configuration.
Conversely, when both are on the same network, no external servers are used to control the camera and access the video stream.

\begin{figure}[ht]
 \centering
 \includegraphics[scale=0.3]{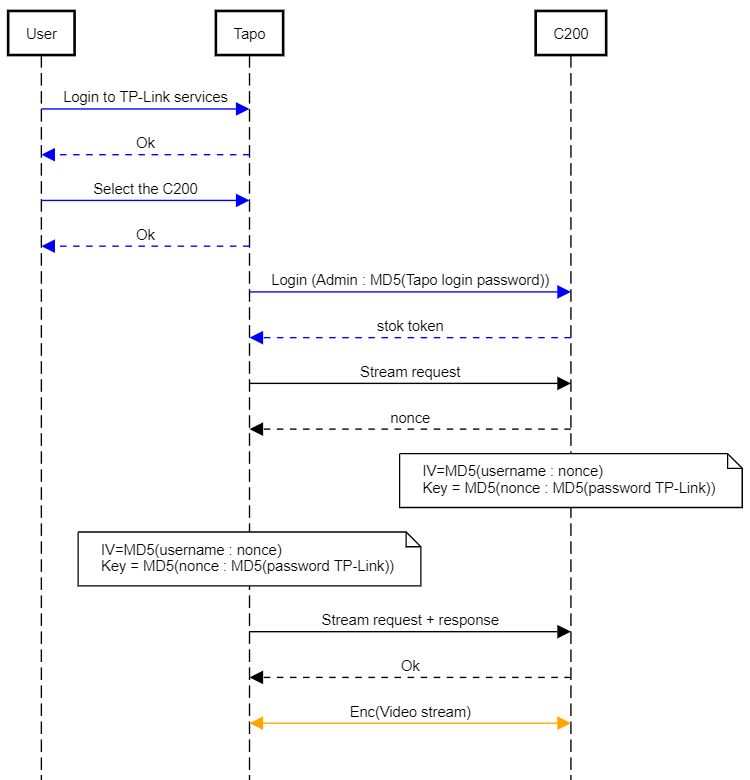}
 \begin{flushleft}
 \fcolorbox{black}{black}{\rule{0pt}{4pt}\rule{4pt}{0pt}}\quad Communication in plaintext.\newline
 \fcolorbox{black}{blue}{\rule{0pt}{4pt}\rule{4pt}{0pt}}\quad Communication over TLS channel.\newline
 \fcolorbox{black}{orange}{\rule{0pt}{4pt}\rule{4pt}{0pt}}\quad Stream video Encrypted with AES key
 \end{flushleft}
 \caption{Sequence diagram for using the IP Camera with the proprietary Tapo application}
 \label{ctrltapo}
\end{figure}

It is important to note that the flow for control and editing differs substantially from that for video streaming.
The diagram in Figure \ref{ctrltapo} describes the sequence of operations that allow the user to access the video stream through the Tapo application. After choosing one of the initialised devices, the Tapo application logs into the Tapo C200, obtaining the token \emph{stok} needed to perform the control and modification operations.
The \emph{stok} is a token generated during authentication by the Tapo C200 which, after authenticating the user with a username and password, assigns this token to the Tapo application. The purpose of the token is to create a session without having to re-authenticate the application every time the camera settings are changed. The various requests generated by the application carry the \emph{stok} obtained in the last authentication phase. In this way the Tapo C200 only accepts requests from authenticated users.

Following the user's request to access the video stream the Tapo application and the Tapo C200 coordinate, in particular, the Tapo C200 sends to the application a \textit{nonce} necessary for the construction of the symmetric key.
At this point both parties calculate the initialisation vector and AES key in order to encrypt the video stream using AES in CBC mode.
Once the key has been calculated, the Tapo application authenticates itself to the Tapo C200 by providing an \lq\lq response" tag. Finally, the Tapo C200, after verifying the validity of the response, starts the multimedia streaming session by encrypting it with the agreed key.

\begin{figure}[ht]
 \centering
 \includegraphics[scale=0.32]{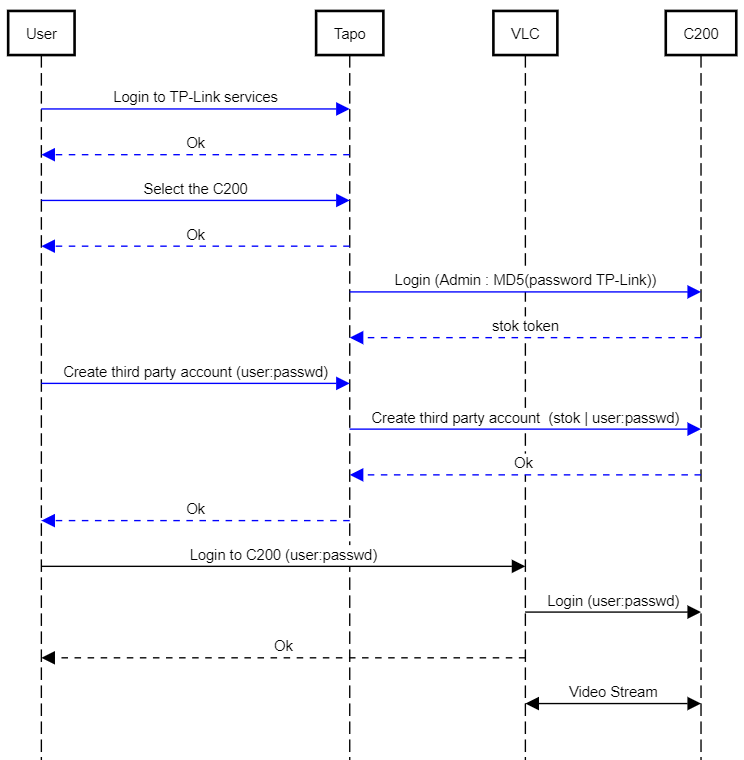}
 \begin{flushleft}
 \fcolorbox{black}{black}{\rule{0pt}{4pt}\rule{4pt}{0pt}}\quad Communication in plaintext.\newline
 \fcolorbox{black}{blue}{\rule{0pt}{4pt}\rule{4pt}{0pt}}\quad Communication over TLS channel
 \end{flushleft}
 \caption{Sequence diagram for using the IP Camera with third party software}
 \label{diagterz}
\end{figure}

From the diagram in Figure~\ref{diagterz} we can notice how the integration of the Tapo C200 with third-party software is handled. Using the Tapo application, we first need to access the device in question and change its settings so that we can create a new user for third-party software.
After that, we can use the Tapo C200 with the previously mentioned software iSpy and VLC, which log in and start the free-to-air media streaming session with the camera.

In order to be able to use third-party applications, it was necessary to correctly configure the already initialised Tapo C200 by creating a special user on the camera using the Tapo application. Therefore, since this was a change to the Tapo C200 settings, the request generated by the Tapo application was accompanied by the \emph{stok}. 

To access the resource, the Linux machine with VLC and iSpy installed was used, i.e. the machine (e) of the testbed. Using Ettercap and Wireshark installed on machine (d) instead, a MiTM attack was carried out in order to analyse the traffic.
From machine (e) it was possible to access the video stream using both players.

\begin{itemize}
 \item On VLC, select the \lq\lq network resources" section of the player and enter:\\
 \url{rtsp://username:password@<tapoc200address>/stream/1}
 \item On iSpy, configure the ONVIF S profile to URI:\\ \url{http://username:password@<tapoc200address>/onvif/device\_service}
\end{itemize}

Since ONVIF's S profile does not add any functionality to the video stream, no differences emerged from the analysis of the traffic generated for the video stream of the two players. In particular, the packets analysed showed that, after accessing through the specific URI for ONVIF devices, iSpy uses RTSP in the exact way used by VLC. The diagram shows how the Tapo C200 is integrated with third-party software. Using the Tapo application, we first need to access the device in question and change its settings so that we can create a new user account for third-party software.
In summary, the Tapo C200 can then be used with the previously mentioned software iSpy and VLC, which log in and start the unencrypted multimedia streaming session with the camera.

\section{Vulnerability Assessment}~\label{sec:VA}
The vulnerability assessment activity carried out is divided into two parts, one concerning proprietary services and the other concerning third party video streams.
\subsection{Proprietary services analysis}
Security in terms of the confidentiality of the video stream when the Tapo application and Tapo C200 are not on the same network, is entrusted to the channel itself on which it takes place. Using SSL/TLS on port 443, all data exchanged between the Tapo C200, the TP-Link server and the Tapo application is properly encrypted.
In addition, the Tapo application and the Tapo C200 use SSL pinning~\cite{pinning}, meaning that if information needs to be shared with the TP-Link server, no SSL/TLS session is established towards any entity other than a TP-Link server whose SSL/TLS certificate they already have by default.
However, despite the use of SSL/TLS, it was found that the traffic is easily detected, particularly with regard to movement notifications.
The motion detection function follows the steps:

\begin{enumerate}
  \item The Tapo C200 detects movement and sends notification to the TP-Link server;
  \item The server receives the notification and sends an alert to all devices connected to the account associated with the camera;
  \item The application receives the message and generates a notification for the user.
\end{enumerate}
Our study found that the size of the messages generated was always the same, 523 bytes. By using the Tapo C200 as an oracle, the attacker may be able to discern movement notifications based on the size of the messages, without intervening in the cryptographic scheme and deducing, for example, when the victim is home and when no one is home. In addition, by filtering and blocking only those types of messages, the attacker could undermine the availability of the notification service by deceiving the victim that there is no movement in the area covered by the camera.

\subsection{Third party video stream analysis}
The use of third-party players for multimedia streaming presents a high level criticality, both as regards the ONVIF service on port 2020 and RTSP on port 554.
Since the ONVIF profile S is limited to providing an interface for the use of RTSP, both approaches seen for video streaming do not have any system to guarantee the confidentiality of the multimedia stream. In fact, after configuring the Tapo C200 to use these systems, whenever a request is made, the video stream is sent in clear text following the H264 encoding.
This approach is critical, especially with regard to the confidentiality of the data as it could allow the attacker to intercept the stream and decode it, thus obtaining the original and reproducible video stream.

\section{Penetration Testing}~\label{sec:PT}
\subsection{Camera DoS} \label{sec:camerados}
\begin{figure}[ht]
 \centering
 \includegraphics[scale=0.3]{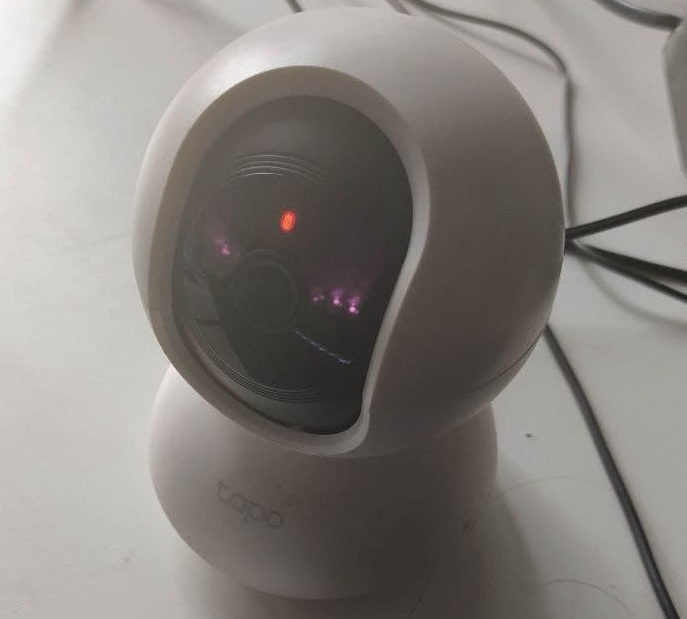}
 \caption{Crash of the Tapo C200 camera}
 \label{fig:crashed_camera}
\end{figure}

A Denial of Service (DoS) is a malfunction of a device due to a computer attack. Generally, such attacks aim to undermine the availability of a service by overloading the system and making an arbitrarily large number of requests, sometimes in a distributed manner (Distributed Denial of Services or DDoS). In the case of the Tapo C200, a DoS attack was relatively simple. In fact, an intensive scan by Nessus caused the device to crash, as can be seen in Figure~\ref{fig:crashed_camera}, causing it to stop and then reboot. This demonstrates how an attacker on the same network as the device can seriously compromise the availability of the service.

\begin{figure}[ht]
 \centering
 \includegraphics[scale=0.5]{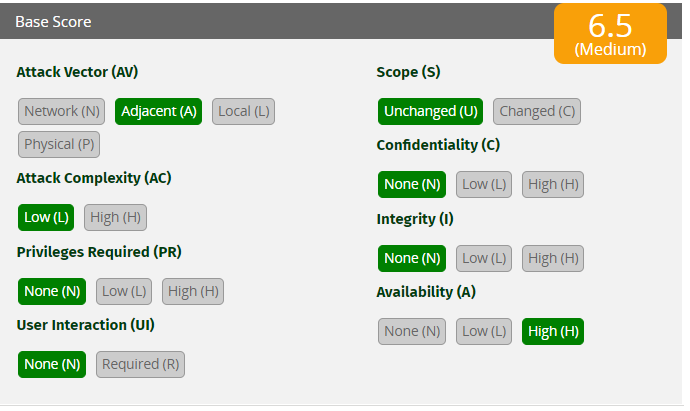}
 \caption{DoS score}
 \label{fig:DoSScore}
\end{figure}

Figure~\ref{fig:DoSScore} shows the CVSS score of this vulnerability,
which has been classified as Medium, with a value of 6.5.

\subsection{Video Eavesdropping} \label{sec:privacyattack}
\begin{figure}[ht]
\centering
 \includegraphics[scale=0.4]{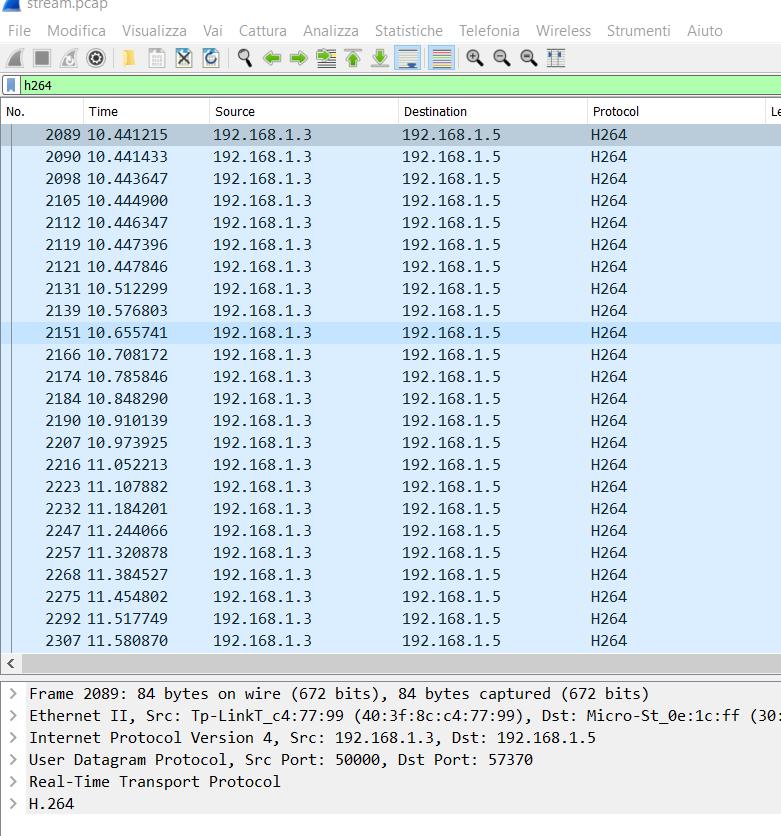}
 \caption{Packages H264}
 \label{extractor}
\end{figure}

\begin{figure}[ht]
\centering
 \includegraphics[scale=0.5]{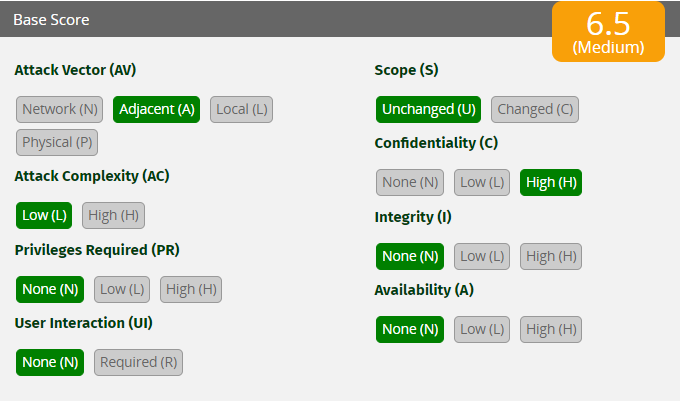}
 \caption{Video eavesdropping score}
 \label{fig:leakscore}
\end{figure}

To demonstrate how an attacker could obtain the multimedia stream transmitted using third-party software, an experiment was conducted using the machine on which the multimedia players were installed, that is the machine (e), and starting a multimedia session with the Tapo C200. At the same time, machine (d) was used again in MiTM attack mode to intercept the traffic exchanged between the player and the Tapo C200.
By analysing the intercepted traffic with Wireshark it was possible to use the H264 extractor tool~\ref{extractor}, the tool was able to correctly interpret and reconstruct the entire intercepted multimedia streaming session. The output of the tool, produce the reproducible video of the intercepted session.
Finally, Figure~\ref{fig:leakscore} shows the CVSS score of this vulnerability, which was classified as Medium, with a value of 6.5

\subsection{Motion Oracle} \label{sec:motionoracle}
\begin{figure}[ht]
\centering
 \includegraphics[width=\linewidth]{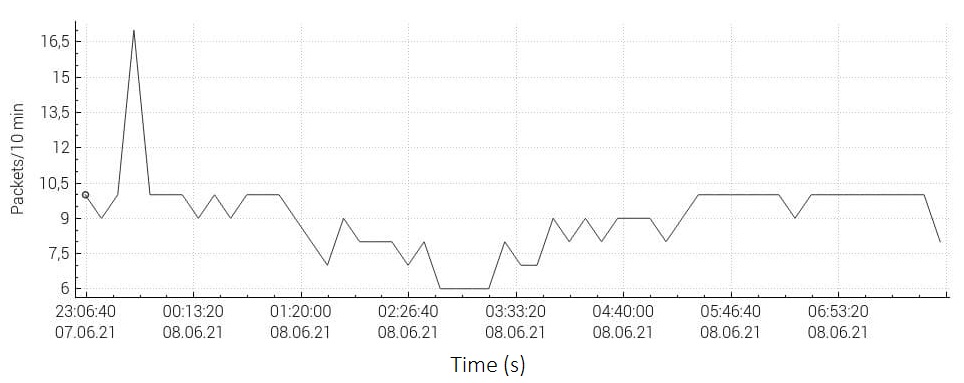}
 \caption{Graph of transit packet traffic on 7-8 June 2021}
 \label{graftraf}
\end{figure}

\begin{figure}[ht]
\centering
 \includegraphics[width=\linewidth]{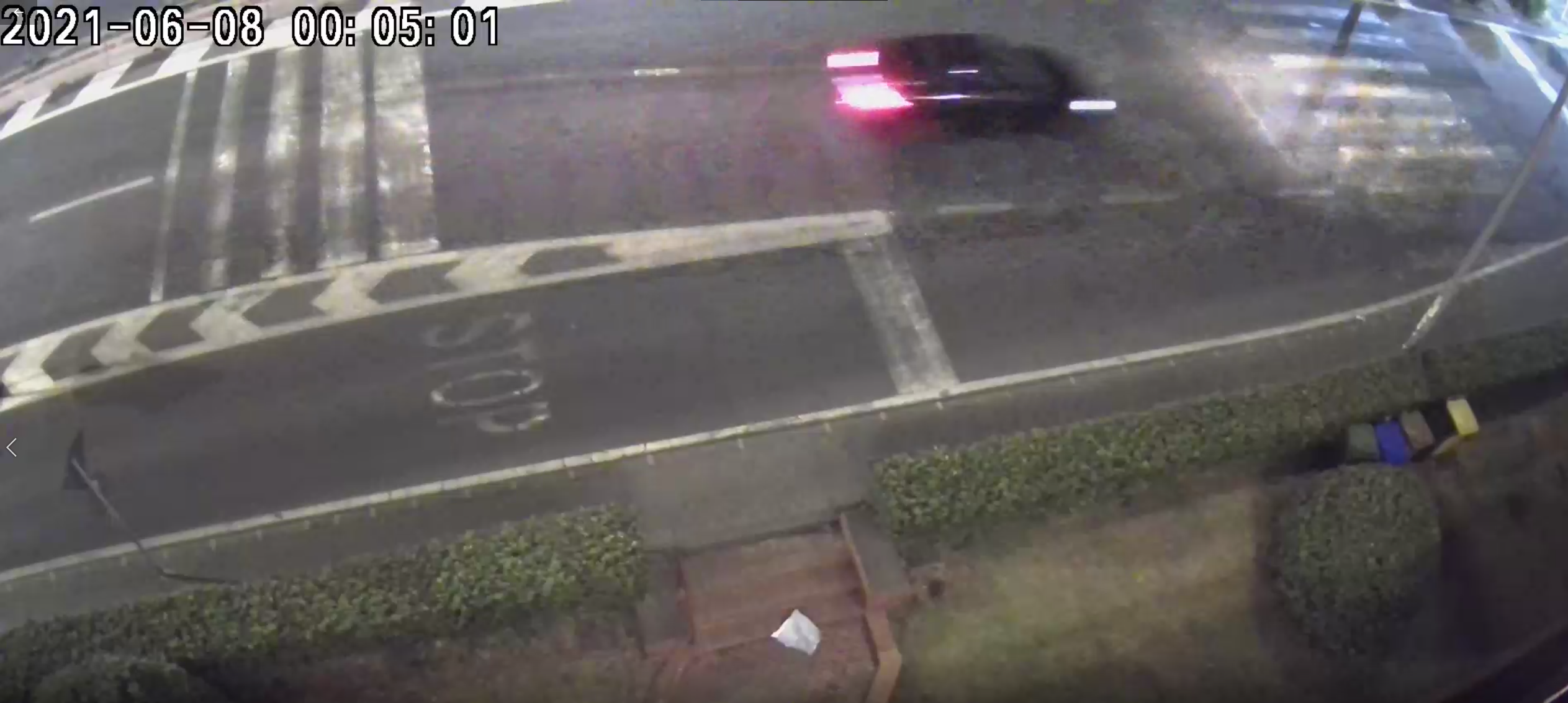}
 \caption{Movement recorded at 00:05}
 \label{immovi}
\end{figure}

To demonstrate how an attacker could use the Tapo C200 as an oracle, a special experiment was conducted.
Throughout the night, the Tapo C200 was used to record a public street. Once the traffic was recorded, outgoing packets from the Tapo C200 were filtered on the base:
\begin{itemize}
  \item Of the protocol used, by inserting the SSL/TLS filter
  \item Of the frame length, by inserting the filter \textit{frame.length==523}
\end{itemize}
Once the packets were obtained, we constructed a graph shown in Figure~\ref{graftraf}. From the graph we can see the number of SSL/TLS packets of 523 bytes captured at ten minute intervals. In addition, it can be seen that starting at 23:00, the amount of packets begins to decrease until about 3:00. From 3:30 it is possible to see an increase in packets reaching a maximum around 7:00. The trend of the curve is consistent with what could be expected and shows that the influx of cars decreases during the night and increases in the morning.
By comparing the graph with the video recorded on the Tapo C200's SD card, it was found that there was a complete match between captured packets and recorded movements. Observing the recorded video, it was found that all the actual movements are shown in the graph. For example, a frame from about 00:05 is shown in Figure~\ref{immovi}, where a car can be seen in transit.

\begin{figure}[ht]
\centering
 \includegraphics[scale=0.5]{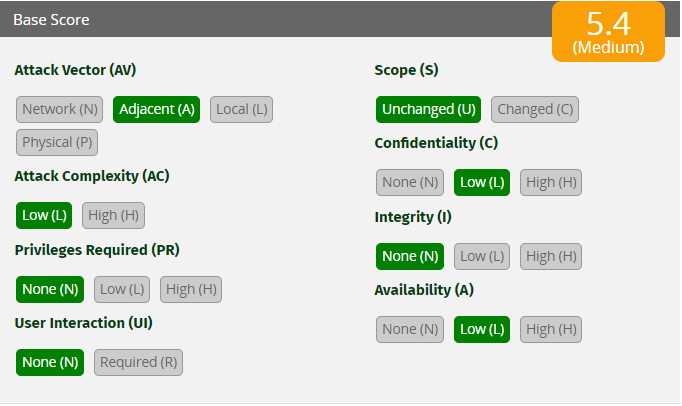}
 \caption{Motion Oracle score}
 \label{scoremo}
\end{figure}

This demonstrates how an attacker on the same network as the Tapo C200 can precisely detect traffic related to the Tapo C200's motion detection service, traffic that could be used to deduce information about the victim by monitoring the frequency of these packets.
Furthermore, this experiment demonstrates how the attacker can trivially filter out such packets and deceive the victim.
Figure~\ref{scoremo} shows the CVSS score of this vulnerability, which was classified as \emph{Medium}, with a value of 5.4.

\section{Countermeasures}\label{sec:countermeasures}
Our work proposes a set of countermeasures developed to mitigate the vulnerability of video eavesdropping. The Raspberry Pi 4 Model B was used for this purpose.
The idea behind the proposed solution to guarantee the confidentiality of the video stream in the third party scenario is to use the Raspberry Pi 4 Model B as an access point of the Tapo C200 in order to modify the traffic in transit by applying encryption.

\subsection{Raspberry Pi into the testbed}
\begin{figure}[ht]
\centering
 \includegraphics[scale=0.35]{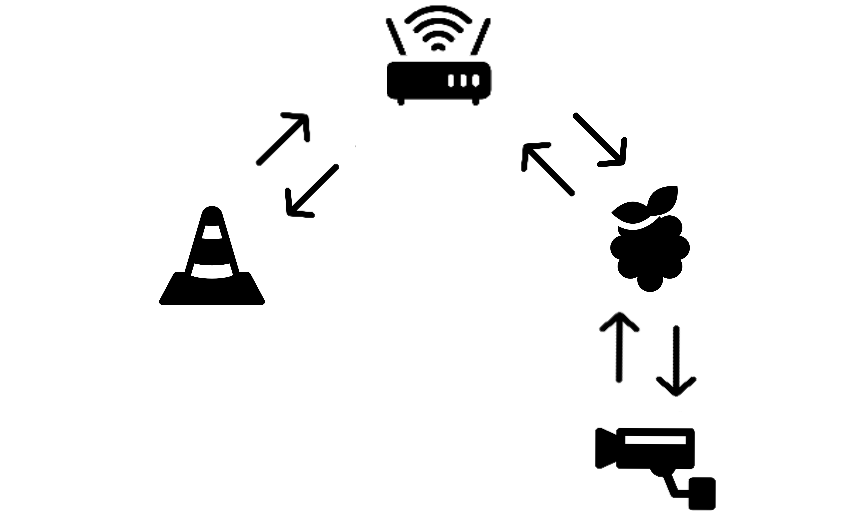}
 \caption{Raspberry Pi in the network infrastructure}
 \label{testbed2}
\end{figure}

The Raspberry Pi 4 Model B has a network card with a wired and a wireless interface. Thanks to this feature it was possible to connect the Raspberry Pi 4 Model B to the testbed modem while exposing a Wi-Fi network to which the Tapo C200 could be connected, as shown in Figure \ref{testbed2}.

In order to build the network infrastructure described above, the Raspberry Pi 4 Model B had to be configured to act as an access point for the Tapo C200 on behalf of the modem~\cite{settingrasp}. Specifically, it was necessary to install the hostapd~\cite{hostpad} tool.
This module, available for Linux machines, allows normal network interfaces to be transformed into Access Points. In addition, it was necessary to modify the dhcpd configuration of the Raspberry Pi 4 Model B.

\subsection{Traffic encryption}
To encrypt network traffic we configured IPtables on the Raspberry Pi 4 Model B, the policy adopted is shown in Listing~\ref{code:IPtablesEnc}:

\begin{lstlisting}[caption={Configuration of IPtables for encryption},  label={code:IPtablesEnc}]
iptables - A FORWARD - m 
--physdev-is-in wlan0 -p udp
-s TAPOC200_ADDRESS_IPV4
-j NFQUEUE --queue num 1

iptables -A INPUT-A FORWARD -m
--physdev-is-in wlan0 -f -j DROP
\end{lstlisting}

In this way it was possible to intercept packets in transit (FORWARD) from the network interface (-m --physdev-is-in wlan0) transported with UDP (-p udp) coming from the Tapo C200 (-s TAPOC200\_ADDRESS\_IPV4) and send them on queue 1.
Thanks to the second rule, it is possible to eliminate all possible fragments of the packets in transit, in order to avoid sending partial packets.
At this point, we have built a script to encrypt the packets queued by the firewall. The script takes care of retrieving packets from the queue, extracting the payload, encrypting it and sending it back to the recipient's address, all the script code is shown in the Listing~\ref{code:encryptscript}.

\begin{lstlisting}[caption={Encryption script}, label={code:encryptscript}]
def encrypt(packet):
  cipher_suite = Fernet(key)
  encoded_text = cipher_suite.encrypt
  (packet.get_payload()[28:])
 
  pkt = IP(packet.get_payload())
  UDP_IP = pkt[IP].dst
  UDP_PORT = pkt[UDP].dport
  MESSAGE = encoded_text
  
  sock = socket.socket(socket.AF_INET,
  socket.SOCK_DGRAM)
  sock.sendto(MESSAGE, (UDP_IP, UDP_PORT))

  packet.drop()
  
\end{lstlisting}

\subsection{Traffic decryption}
A configuration of IPtables was also developed for the client with the third-party software running, the policy adopted is shown in Listing~\ref{code:IPtablesDec}:

\begin{lstlisting}[caption={Configuration of IPtables for decryption}, label={code:IPtablesDec}]
iptables -A INPUT -p udp
-s RASPBERRY_ADDRESS_IPV4
-j NFQUEUE --queue num 2 
\end{lstlisting}

Through this solution it is possible to intercept incoming packets (INPUT) transported with UDP (-p udp) coming from the Raspberry Pi 4 Model B (-s RASPBERRY \- ADDRESS \_ IPV4) and send them on queue 2.

The script for decrypting the packets queued by the firewall is shown in Listing~\ref{code:decryptscript}.
More specifically, the script takes care of fetching packets from the queue, extracting the payload and decrypting it before accepting it into the system.

\begin{lstlisting}[caption={Decryption script}, label={code:decryptscript}]
def decrypt(packet):
  cipher_suite = Fernet(key)
  decoded_text = cipher_suite.decrypt
  (packet.get_payload()[28:])
 
  pkt = IP(packet.get_payload())
  UDP_IP = pkt[IP].dst
  UDP_PORT = pkt[UDP].dport
  MESSAGE = decoded_text
  
  sock = socket.socket(socket.AF_INET,
  socket.SOCK_DGRAM)
  sock.sendto(MESSAGE, (UDP_IP, UDP_PORT))
  
  packet.drop()

\end{lstlisting}

\subsection{Proposed solution in the third-party scenario}

\begin{figure}[ht]
\centering
 \includegraphics[scale=0.3]{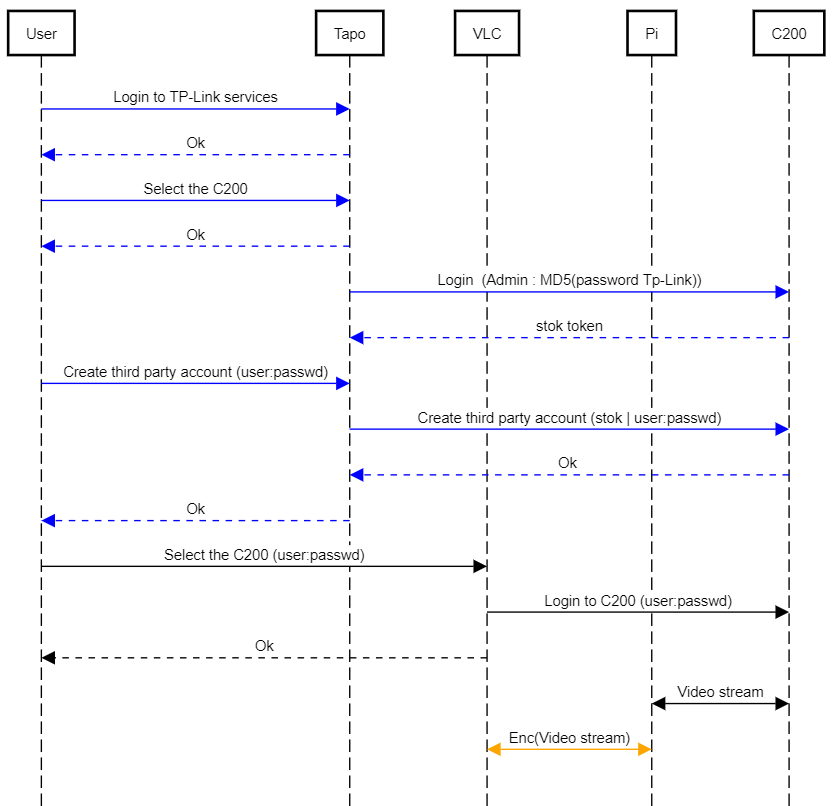}
 \begin{flushleft}
 \fcolorbox{black}{black}{\rule{0pt}{4pt}\rule{4pt}{0pt}}\quad Communication in plaintext\newline
 \fcolorbox{black}{blue}{\rule{0pt}{4pt}\rule{4pt}{0pt}}\quad Communication over the TLS channel\newline
 \fcolorbox{black}{orange}{\rule{0pt}{4pt}\rule{4pt}{0pt}}\quad Video stream encrypted with symmetrical AES key
 \end{flushleft}
 \caption{Sequence diagram of the use of the IP Camera with Raspberry Pi in the network infrastructure}
 \label{utipi}
\end{figure}

The configuration of the countermeasures just seen, although it modifies the initial testbed, has no impact on the actions the user has to take to initialise the Tapo C200 and its use.
A sequence diagram of how the Tapo C200 works with third-party software is shown in Figure~\ref{utipi}.

\begin{figure}[ht]
\centering
 \includegraphics[scale=0.5]{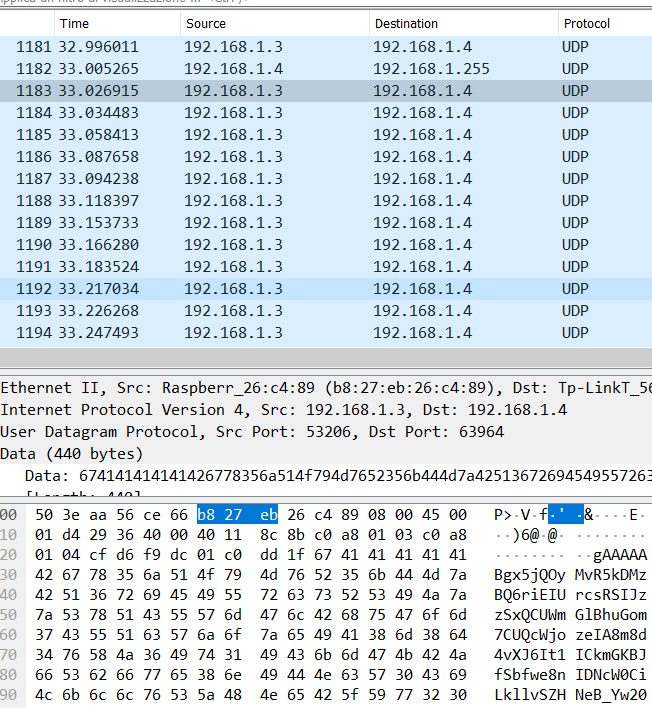} 
 \caption{Encrypted network traffic}
 \label{trafficcif}
\end{figure}

It can be seen that in the use phase the presence of the Raspberry Pi 4 Model B is totally transparent to the user. The routing of the video stream is managed by the devices themselves, so the user can use the service exactly as in the initial scenario. Moreover, thanks to the proposed countermeasure, we can observe that the video stream exchanged on the home network between the camera and the third party application is encrypted.
In order to demonstrate the effectiveness of the countermeasure, the experiment was carried out again, but in this case the execution of the H264 tool failed. The intercepted traffic, shown in Figure~\ref{trafficcif}, was unintelligible and the execution of the H264 extractor tool produced no results.

\section{Conclusions}\label{sec:conclusions}
IP cameras such as the Tapo C200 are very useful, widespread and definitely within everyone's reach. With the experiments conducted in this manuscript we have analysed the security of these devices also from a privacy point of view.
In order to fully understand what Tapo's vulnerabilities might be, a lot of reverse engineering work had to be done using various technologies.
This allowed us to define a detailed description of the ceremony that the Tapo C200 performs during its operation, among these operations we consider: the initialisation of the device, the use of the device through the proprietary systems of TP-Link and the integration of such device with third-party technologies.

Subsequently, the vulnerability assessment was carried out, from which it was found that although video streaming through the use of the Tapo application enjoys certain guarantees, the streaming service through the use of third-party applications is devoid of encryption mechanisms and therefore vulnerable.

Furthermore, it was highlighted that without intervening on the cryptographic systems used, one of the features offered by the camera, namely motion detection, can be a vector to profile and deceive the victim. It has been demonstrated how an attacker could exploit these vulnerabilities through penetration testing. The trial highlighted the Tapo C200's poor resilience to DoS attacks.
Therefore, it was highlighted that by applying a filter on the packets leaving the Tapo C200 on the basis of the protocol and the size, the motion detection service behaves like an oracle for the attacker.
By exploiting the mechanisms in question, the attacker could deduce the victim's habits and surgically inhibit the serve.
Furthermore, it was highlighted that thanks to the use of the H264 extractor tool, the attacker is able to intercept a video stream and convert it back into a playable video. The average of the Base scores of these vulnerabilities is 6.14, so the risk level associated with the Tapo C200 is Medium.

Finally, with regard to the critical issues encountered with the use of third-party technologies, a countermeasure was proposed using the Raspberry Pi 4 Model B. This countermeasure uses the Raspberry as an extension of the Tapo C200, capable of applying encryption to transmissions and guaranteeing confidentiality in a completely transparent way to the user. 

Our experiments taught us that, although the security measures in place in the Tapo C200 are in line with the state of the art, they have not yet been tailored to all possible use scenarios, a finding that we plan to validate in the future over other common IoT devices.

\bibliographystyle{abbrv}
\bibliography{biblio}

\end{document}